\documentclass[aps,pra,twocolumn,footinbib]{revtex4-1}
\usepackage{graphicx}
\usepackage{amsmath,amsthm,amssymb,mathtools}
\usepackage{color}
\usepackage[none]{hyphenat}

\usepackage[pdftex,colorlinks=true,linkcolor=Cerulean,citecolor=Cerulean,urlcolor=Cerulean]{hyperref}
\hypersetup{linkcolor=blue,citecolor=blue,urlcolor=blue}
\hypersetup
{
  pdfauthor={J. Enrique V\'azquez-Lozano \& Alejandro Mart\'inez},
  pdfsubject={Nanophotonics Technology Center -- Universitat Polit\`ecnica de Val\`encia},
  pdftitle={Classical Emergence of Intrinsic Spin-Orbit Interaction of Light at the Nanoscale},
  pdfkeywords={spin-orbit interaction, factorizability (or separability) condition, nanophotonics}
}

\begin{document}
\title{Classical emergence of intrinsic spin-orbit interaction of light at the nanoscale}
\author{J. Enrique V\'azquez\,-Lozano and Alejandro Mart\'inez}
\affiliation{
Nanophotonics Technology Center, Universitat Polit\`ecnica de Val\`encia, Camino de Vera s/n, 46022 Valencia, Spain}

\date{\today}
\begin {abstract} 
Traditionally, in macroscopic geometrical optics intrinsic polarization and spatial degrees of freedom of light can be treated independently. However, at the subwavelength scale these properties appear to be coupled together, giving rise to the spin-orbit interaction (SOI) of light. In this work we address theoretically the classical emergence of the optical SOI at the nanoscale. By means of a full-vector analysis involving spherical vector waves we show that the spin-orbit factorizability condition, accounting the mutual influence between the amplitude (spin) and phase (orbit), is fulfilled only in the far-field limit. On the other side, in the near-field region, an additional relative phase introduces an extra term that hinders the factorization and reveals an intricate dynamical behavior according to the SOI regime. As a result, we find a suitable theoretical framework able to capture analytically the main features of intrinsic SOI of light.  Besides allowing for a better understanding into the mechanism leading to its classical emergence at the nanoscale, our approach may be useful in order to design experimental setups that enhance the response of SOI-based effects.
\end{abstract}
\pacs{}
\maketitle
\sloppy

% % % % % % % % % % % % % % % % % % % % % % % % % % % % % % % % % % % % % % % % % % % % % % % % % % % % % % % % % % % % % % % % % % % % % % % %
% INTRODUCTION
% % % % % % % % % % % % % % % % % % % % % % % % % % % % % % % % % % % % % % % % % % % % % % % % % % % % % % % % % % % % % % % % % % % % % % % %
\section{\label{sec:intro}Introduction}
% GENERAL FRAMEWORK OF SOI
Spin-orbit interaction (SOI) comprises a broad class of effects very well-known in the branches of atomic and solid state physics \cite{Mathur1991,Rashba2006}. Roughly speaking, such phenomena involve charged particles moving within a region where there is an electric field, e.g., that originated by the atomic nuclei or by the asymmetry in the confinement potential of electrons in heterostructures. In these contexts, SOI can be conceived as an effective phenomenon of relativistic nature wherein the motion of the particle is coupled with its spin \cite{Akhiezer1965}. The importance of this interaction is noteworthy since it has allowed to explain the fine structure energy corrections of hydrogen-like atoms. Nonetheless, even more important has been the occurrence of SOI in solids, paving the way to the area of spintronics \cite{Wolf2001}. 

\vspace{0.265cm}
% SOI OF LIGHT: FUNDAMENTAL DEFINITION
The extension of SOI to optics is attributed to the seminal work by Liberman and Zel'dovich \cite{Liberman1992}. Their approach is based on the conservation of the state of polarization (SoP) when light propagation is subjected to bending and/or twisting in an optically inhomogeneous medium. Under this scheme, they introduced the optical SOI as the mutual interaction between the SoP (spin) and the propagation process (orbit). This coupling can be simply characterized in a mathematical way by means of the so-called \textit{factorizability (or separability) condition}, which accounts for the mutual influence between the amplitude and the phase of light.

\vspace{0.265cm}
% SOI OF LIGHT : FIRST DEFINITION (PRAGMATIC POINT-OF-VIEW)
Akin to mechanical systems, light possesses a set of dynamical properties such as energy, linear momentum, and angular momentum among others \cite{Belinfante1940}. Due to the vector character of the electromagnetic fields two types of rotations can be distinguished \cite{Birula2011,Andrews2013}, giving rise to the corresponding contributions termed as spin angular momentum (SAM) \cite{Poynting1909} and orbital angular momentum (OAM) \cite{Allen1992}, respectively. Whereas OAM is related to the spatial distribution and propagation of the optical field, SAM is generally determined by the SoP \cite{Bliokh2013,Barnett2016}. Notice that, from a quantum approach, the correspondence principle states that each of the two possible spin states of photons can be identified with the corresponding right- and left-handed circular polarization. This rule only holds for the usual longitudinal SAM, closely linked to the plane wave representation. Still, this picture is in sharp contrast with the transverse SAM, which is characteristic of evanescent as well as structured optical fields \cite{Aiello2015b,Bliokh2012,Bekshaev2015}. Taking into account the above dynamical quantities, from a pragmatic point of view, the optical SOI is commonly understood as the interplay and mutual conversion between the different types of angular momenta \cite{Vesperinas2015,Abdulkareem2016}. However, this definition only emphasizes into the effects, neglecting its fundamental appearance and leading to a certain controversy related with the proper way in which must be performed the separation of the total angular momentum into the spin and orbital contributions \cite{Nieminen2008,Leader2014,Birula2011,Barnett2016}. In this regard, it is noteworthy to mention that this difficulty may be in turn associated with the so-called \textit{Abraham-Minkowski dilemma}, a long-standing problem concerning with an ambiguity that arises from the real definition of the linear and angular momentum for optical radiation in media. Even though there are a number of influential papers claiming to have solved it (see, e.g., Refs. \cite{Nelson1991,Barnett2010}), this challenging problem still remains as a subject of current interest and debate \cite{Bliokh2017,Silveirinha2017}. Notice also that, in relation with the above example regarding the homonymous phenomenon occurring either in atomic or in solid state physics, the interplay between the different kinds of angular momenta play the same role in this case as the spin-dependent splitting in the electronic energy levels, namely, just as an observable effect but not as the ultimate reason leading to the classical emergence of intrinsic SOI of light.

\vspace{0.21cm}
% SOI OF LIGHT : STATE-OF-THE-ART
In the past few years, SOI of light has been the subject of intense research activity. Huge efforts have been devoted to investigate novel photonic applications and functionalities (for a complete overview on this issue see Ref. \cite{Cardano2015} and references therein), paying little attention to the fundamental theory underlying its origin. In this regard it has only been argued that, since photons are relativistic spin-1 particles, SOI of light is inherent to Maxwell's equations \cite{Bliokh2015a}, arising from the transversality condition \cite{Bliokh2010} and described in terms of the geometric Berry phase formalism \cite{Bliokh2009}. Furthermore, the close relation between SOI and the intrinsic spin Hall effect of light has been extensively studied, both theoretically and experimentally \cite{Onoda2004}. The latter manifests itself as a topological spin-dependent transport of photons taking place in inhomogeneous media as well as in free space, thereby ensuring the conservation of the total angular momentum \cite{ Haefner2009}. Additional spin-related optical phenomena such as the aforementioned transverse spin \cite{Bliokh2014}, topological insulators \cite{Khanikaev2013} or spin-momentum locking \cite{Mechelen2016}, leading to the so-called spin-controlled unidirectional excitation \cite{Fortuno2013}, have been recently demonstrated as manifestations of the quantum spin Hall effect of light \cite{Bliokh2015b}. It is evident the vast and unified body of knowledge that exists around SOI. Nonetheless, as already stated, its occurrence is ultimately justified from elementary effects such as the Rytov-Vladimirskii-Berry rotation, the Imbert-Fedorov transverse shift, or the optical Magnus effect \cite{Liberman1992}. In addition, it is important to stress that the subwavelength character of the optical SOI is a prescription that, although widely assumed and confirmed both experimentally and numerically by means of rather qualitative arguments stemming from its effects, to the best of our knowledge, still remains without any reliable analytical demonstration that supports it.

% MAIN CLAIM
In this paper we aim to provide further understanding into the classical emergence of optical SOI. From a full-vector description based on the multiple-multipole method \cite{Novotny}, we demonstrate analytically that SOI of light is a phenomenon that naturally and necessarily come into play at the subwavelength scale, even in homogeneous media. Indeed, by using the formalism of vector spherical wave functions (VSWFs) \cite{Jackson} in combination with the above-mentioned factorizability condition, we find an additional relative phase that introduces an extra term enclosing the main features of SOI, i.e., it prevents the amplitude-phase separability, but solely in the near-field region. Although it seems a somewhat trivial standpoint, this is certainly the key point in order to demonstrate the universal occurrence of optical SOI at the nanoscale. Of course, this approach satisfy the overall prescriptions underlying the SOI of light, i.e., it is implicit in the Maxwell's equations, and is ultimately related to the transversality condition of the electromagnetic fields. Importantly, our results also allow us to identify accurately the region wherein SOI-based phenomena naturally emerge. Therefore, besides providing a more fundamental definition for the near-field region in terms of the factorizability condition, they may be used to facilitate or improve the setups for the experimental observation of SOI-based effects.

% % % % % % % % % % % % % % % % % % % % % % % % % % % % % % % % % % % % % % % % % % % % % % % % % % % % % % % % % % % % % % % % % % % % % % % %
% OVERVIEW OF FULL-VECTOR WAVES
% % % % % % % % % % % % % % % % % % % % % % % % % % % % % % % % % % % % % % % % % % % % % % % % % % % % % % % % % % % % % % % % % % % % % % % %
\section{\label{sec:VSWFs}Overview of full-vector waves}
To start with, let us consider an arbitrary electromagnetic wave which propagates in a homogeneous medium. By means of the angular spectrum representation, this field can be expressed as a superposition of elementary plane waves, each having a well-defined SoP, constant over the whole space:
\begin{equation}
{\bf E}({\bf r})=\left(\alpha{\bf e}_1+\beta{\bf e}_2\right)E_0({\bf r}) e^{i\phi({\bf r})},
\label{Eseparable}
\end{equation}
where $\alpha$ and $\beta$ are arbitrary complex constants describing the normalized SoP ($|\alpha|^2+|\beta|^2=1$), ${\bf e}_{1,2}$ are two orthogonal unit vectors, $E_0$ is the scalar field profile, and $\phi$ is the phase distribution. As pointed out above, following Ref. \cite{Liberman1992}, SOI of light is envisioned from a fundamental approach as the mutual influence between the SoP and the phase distribution. Hence, owing to the factorized form of the plane wave in Eq. \eqref{Eseparable}, the mutual interaction between the SoP and the phase vanishes, avoiding the occurrence of SOI. This is the usual picture in macroscopic geometrical optics \cite{Wolf}, wherein light is characterized by means of propagating rays which, in turn, can be described as a field expansion into local plane waves. This scalar-like scheme can also be extended to the zeroth-order of the paraxial approximation \cite{Liberman1992,Bliokh2009}. Nevertheless, at the nanoscale, near or beyond the diffraction limit, this usual treatment based on the plane-like waves seems to be pretty naive. Furthermore, due to the extraordinary properties of the angular momentum associated to evanescent fields \cite{Bliokh2014}, the understanding of SOI-based effects for such kinds of fields deserves a special approach (for further details on this issue see Ref. \cite{Fernandez2016}).

Close to the sources, or in the near-field region of the processes wherein light-matter interaction takes place, the spatial field distribution of electromagnetic waves displays complex shapes. Therefore, in order to deal with nontrivially structured optical fields, it become necessary to perform a full-vector wave analysis \cite{Novotny}. Regardless of the spatial distribution, any optical field can be generally expressed as a multipole expansion \cite{Jackson}, i.e., as a proper linear combination of the vector spherical harmonics (VSH). In this way, the electromagnetic field is assumed as the radiated from a point-like source, thus providing a suitable tool to deal with phenomena occurring at the subwavelength scale. This includes the SOI of light as well \cite{Rodriguez2010}, which has been experimentally demonstrated to induce subtle observable effects upon the far-field via imaging systems \cite{Bliokh2011}. Hence, instead of the plane-wave basis, VSWFs seems to be a better choice to accomplish a full description of processes at the nanoscale.

It is well-known that in a source-free, homogeneous, isotropic and linear medium, the time-independent electromagnetic fields can be obtained from the vector Helmholtz wave equation,
\begin{equation}
\nabla\times\left[\nabla\times{\bf \Psi}({\bf r})\right]-k^2 \boldsymbol{\Psi}({\bf r})=0,
\label{Helmholtz}
\end{equation}
where $\boldsymbol{\Psi}({\bf r})$ can be either the electric or magnetic field, $k=n\omega/c$ is the wavenumber, and $n=\sqrt{\varepsilon\mu}$ is the refractive index, being $\varepsilon$ and $\mu$ the corresponding relative permittivity and permeability of the medium. Although there exist several conventions to define the VSH, throughout this work we will follow that given in Ref. \cite{Barrera1985}:
\begin{subequations}
\begin{align}
&{\bf{R}}_{lm}(\varOmega)={\bf e}_r Y_{lm}(\varOmega),\\
&\boldsymbol{\Theta}_{lm}(\varOmega)= N_l r\nabla  Y_{lm}(\varOmega),\\
&\boldsymbol{\Phi}_{lm}(\varOmega)= N_lr\nabla{Y_{lm}(\varOmega)}\times {\bf e}_r,
\end{align}
\end{subequations}
where $N_l=1/\sqrt{l(l+1)}$ is a normalization constant, $Y_{lm}(\varOmega)$ are the scalar spherical harmonics of order $(l,m)$, and $\varOmega\equiv(\theta,\varphi)$ represents the standard angular coordinates (i.e., polar and azimuthal angles, respectively). Taking into account that the VSH form an orthogonal and complete set of basis vectors, any source-free electric (or magnetic) field can be expanded in terms of the VSWFs as follows,
\begin{equation}
{\bf E}({\bf r})=\sum_{l=1}^\infty\sum_{m=-l}^{l}{{\alpha_{lm}{\bf E}^{\rm TE}_{lm}({\bf r})+\beta_{lm}{\bf E}^{\rm TM}_{lm}({\bf r})}},
\label{multipole_exp}
\end{equation}
where $\alpha_{lm}$ and $\beta_{lm}$ are the multipole expansion coefficients (also called beam-shape coefficients), and ${\bf E}^{\rm TE}_{lm}\equiv E_{l}^{(\Phi)}{\bf \Phi}_{lm}$ and ${\bf E}^{\rm TM}_{lm}\equiv E_{l}^{(\Theta)}{\bf \Theta}_{lm}+E_{l}^{\rm{(R)}}{\bf R}_{lm}$ are, respectively, the mutually perpendicular \textit{transverse electric} (TE) and \textit{transverse magnetic} (TM) multipole fields of $(l,m)$-order \cite{footnote1}. It is important to note that, in each of both subsets of solutions, each element verifies individually  the vector Helmholtz equation \eqref{Helmholtz}. Furthermore, the radial dependence of each VSWF is incorporated into the $E_{l}^{(\cdot)}$ functional coefficients and appears separately from the angular coordinates, thereby allowing an independent treatment. In order to determine their specific form we should substitute this last expression \eqref{multipole_exp} into the vector Helmholtz equation \eqref{Helmholtz}. In this manner, it can be demonstrated that the radial distribution is given in terms of the solutions of the spherical Bessel differential equation, which explicitly read as follows \cite{Wolf,Jackson}:
\begin{subequations}
\label{coeffs} 
\begin{eqnarray}
E_{l}^{\rm{(R)}}(r)&=&\frac{f_l(kr)}{N_lkr},\label{coeffa}\\
E_{l}^{(\Theta)}(r)&=&\frac{\left(krf_l(kr)\right)'}{kr},\label{coeffb}\\  
E_{l}^{(\Phi)}(r)&=& f_l(kr),\label{coeffc}
\end{eqnarray}
\end{subequations}
where $f_l(kr)\equiv\left\{j_l(kr),y_l(kr)\right\}$ are the $l$-dependent Bessel-like functions, and the prime denotes differentiation with respect to the dimensionless variable $kr$. Notice that, since there are two independent solutions ($j_l(kr)$ and $y_l(kr)$, being the spherical Bessel functions of the first and second kind, respectively), any linear combination will also be a proper solution. This provides the physical meaning for the radial functions depending on the specific situation. Indeed, if we consider a time-harmonic dependence of the form $e^{-i\omega t}$, in order to describe propagating spherical waves (purely outgoing or incoming) we will use the spherical Hankel functions $h_l^{(\pm)}(kr)=j_l(kr)\pm iy_l(kr)$. On the other hand, singularity-free spherical Bessel functions, $j_l(kr)$, are appropriate functions for representing standing or regular waves. 

% % % % % % % % % % % % % % % % % % % % % % % % % % % % % % % % % % % % % % % % % % % % % % % % % % % % % % % % % % % % % % % % % % % % % % % %
% CLASSICAL EMERGENCE OF OPTICAL SPIN-ORBIT INTERACTION
% % % % % % % % % % % % % % % % % % % % % % % % % % % % % % % % % % % % % % % % % % % % % % % % % % % % % % % % % % % % % % % % % % % % % % % %
\section{\label{sec:SOI}Optical spin-orbit interaction}
\subsection{Intrinsic evolution of the SoP}
\begin{figure}[t!]
\includegraphics[width=0.95\linewidth]{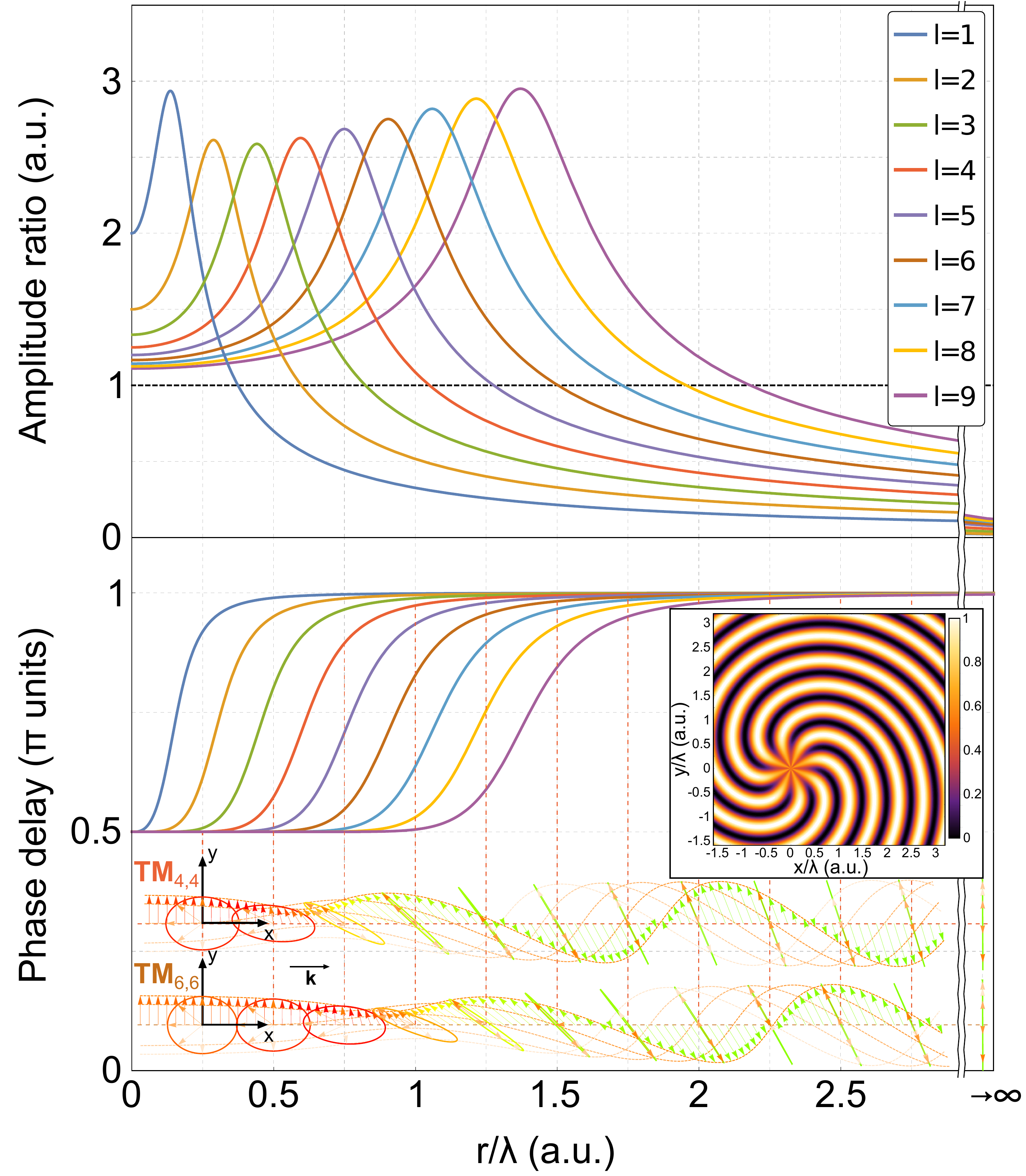} 
\caption{Amplitude ratio (upper panel) and phase delay (lower panel) between the longitudinal and transverse components for different propagating TM multipole fields of $(l,l)$-order with $l\in [1,9]$, over the $xy$-plane. The inset shows the locally normalized instantaneous intensity distribution of the ${\bf E}^{\rm TM}_{4,4}$-mode, whose spatially varying polarization ellipse along the $x$ axis is schematically depicted at the bottom. For comparison, the evolution of the SoP for the ${\bf E}^{\rm TM}_{6,6}$-mode is also plotted. The color coding used in the evolution of the ellipses illustrates the transition from the near- to the far-field zone according to the scale represented in Fig. \ref{fig3}, with the corresponding values for the azimuthal mode order $l$.}
\label{fig1}
\end{figure}
Unlike the aforementioned plane-wave scheme, where the SoP of each field was conserved over the whole space, it can be readily observed that the SoP of the multipole fields is spatially inhomogeneous. This fact is precisely the first hint into the emergence of intrinsic SOI at the nanoscale, even in homogeneous media. We illustrate this idea in Fig. \ref{fig1} by means of the amplitude ratio and the phase delay profiles between the longitudinal ($\propto{\bf R}_{lm}$) and transverse ($\propto {\bf \Theta}_{lm}$) components of different propagating TM-modes of $(l,l)$-order over the $xy$-plane. By considering the electric field contribution, it should be noted that TE-modes are purely transverse, and therefore, the SoP manifests essentially a plane-like behavior, except for the attenuation factor $1/kr$ ensuring the energy conservation. Instead, TM-modes encloses generally the joint action of the longitudinal field component (with a possible transverse SAM), together with the transverse one, owning the main features of SOI (see more details below). The upper panel of Fig. \ref{fig1} shows that, in the near-field zone $|{\bf E}^{\rm TM\rm{(R)}}_{lm}|/|{\bf E}^{\rm TM(\Theta)}_{lm}|>1$, namely, the dominant contribution for these individual $(l,l)$-modes is due to the longitudinal component. This crucial remark, according to which the transverse field component is screened by the longitudinal one, agrees with the already predicted difficulty into the experimental observation of optical SOI \cite{Cardano2015,Bliokh2015a}. On the other hand, the phase delay curves reveal the presence of a relative phase between both components, causing an intricate evolution of the polarization ellipse along the trajectory. Remarkably, the most significant variation takes place in the near-field region, i.e., where the relative phase changes drastically from $\pi/2$ to $\pi$, and coincides with the range for which the amplitude ratio is maximum. As expected, in the far-field limit, the relative phase becomes a constant value and the amplitude ratio goes to zero, thus forcing to the polarization plane to keep it purely orthogonal to the propagation direction. Still, an additional intriguing property is the presence of a minimum in the envelope curve of the amplitude ratio that occurs only for the ${\bf E}^{\rm TM}_{3,3}$-mode. Indeed, in Fig. \ref{fig1} we can see that for each $(l,m)$-order there is an absolute maximum value in the amplitude ratio. Surprisingly, the trend in their magnitudes with respect to the azimuthal mode order $l$ is not trivial, showing a minimum for $l=3$. Therefore, according to the above arguments, this could enable to set an optimal multipole field distribution in order to facilitate the observation of SOI-based effects \cite{Curto2011}. Finally, it should be noted that the regions which we refer to as the near- and far-field zone are ultimately determined by the azimuthal index $l$ (also called topological charge), which is in turn tied to the intrinsic OAM of the corresponding mode \cite{Allen1992}. As it will be shown below, we can find a more accurate definition for these regions via the spin-orbit factorizability condition, thus endowing it with a more fundamental sense and removing the arbitrariness related to the dependence on the distance from the source \cite{Jackson}.

\subsection{Factorizability condition and SOI-term}
\begin{figure}[t!]
	\includegraphics[width=0.95\linewidth]{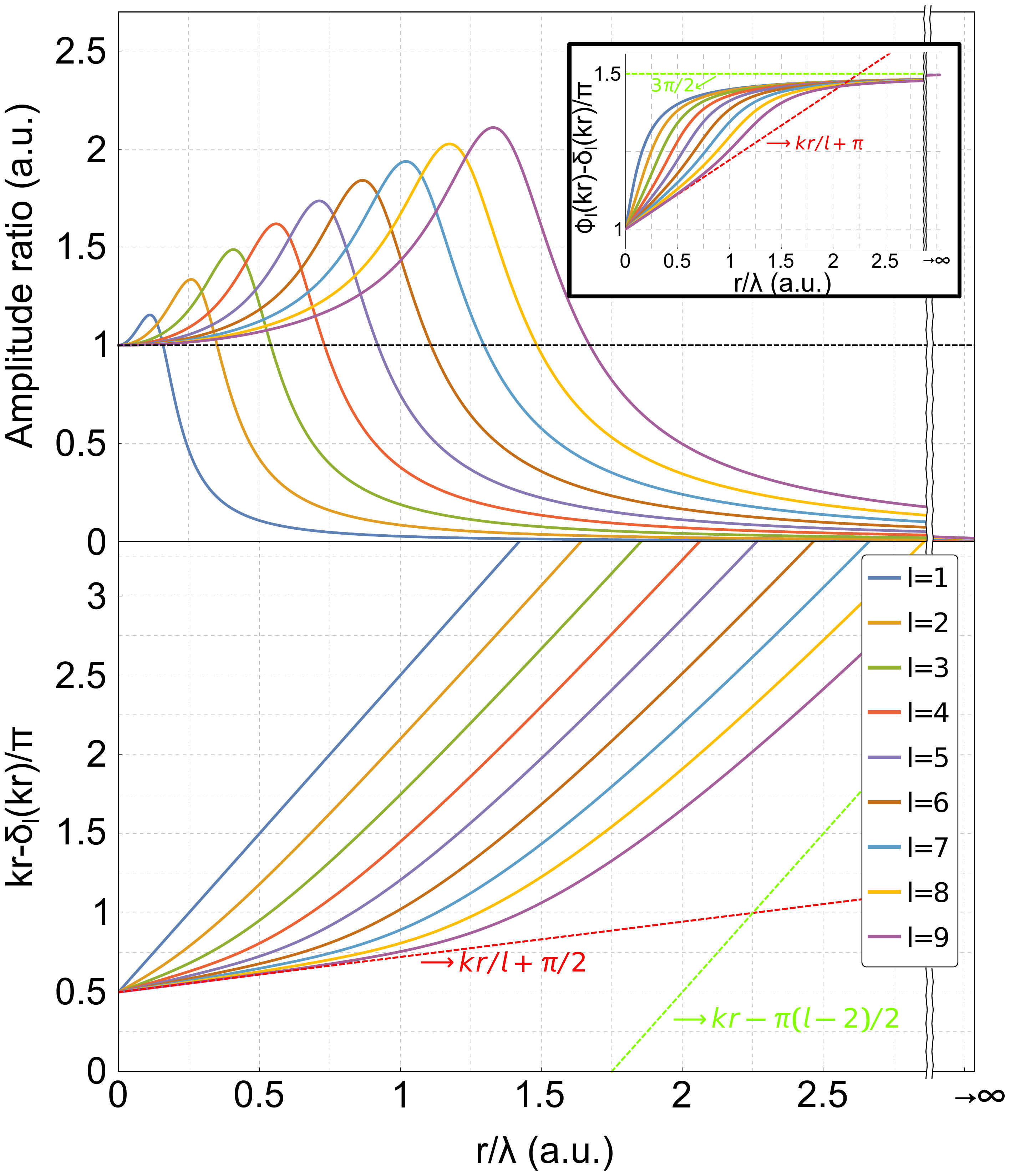} 
	\caption{Main features of the SOI-term $\Delta_l^{(\pm)}$ for different values of $l$. Upper panel displays the amplitude ratio between the SOI-term and the whole $E_l^{(\Theta)}$ function, i.e., $|\Delta_l^{}|/|E_l^{(\Theta)}|$. Panel at the bottom shows the phase distribution of the SOI-term. Dashed lines indicate the asymptotic behavior both in the near- (red) and far-field (green) regions. The inset shows the relative phase $\phi_l(kr)-\delta_l(kr)$.}
	\label{fig2}
\end{figure}
\begin{figure}[t!]
	\includegraphics[width=0.95\linewidth]{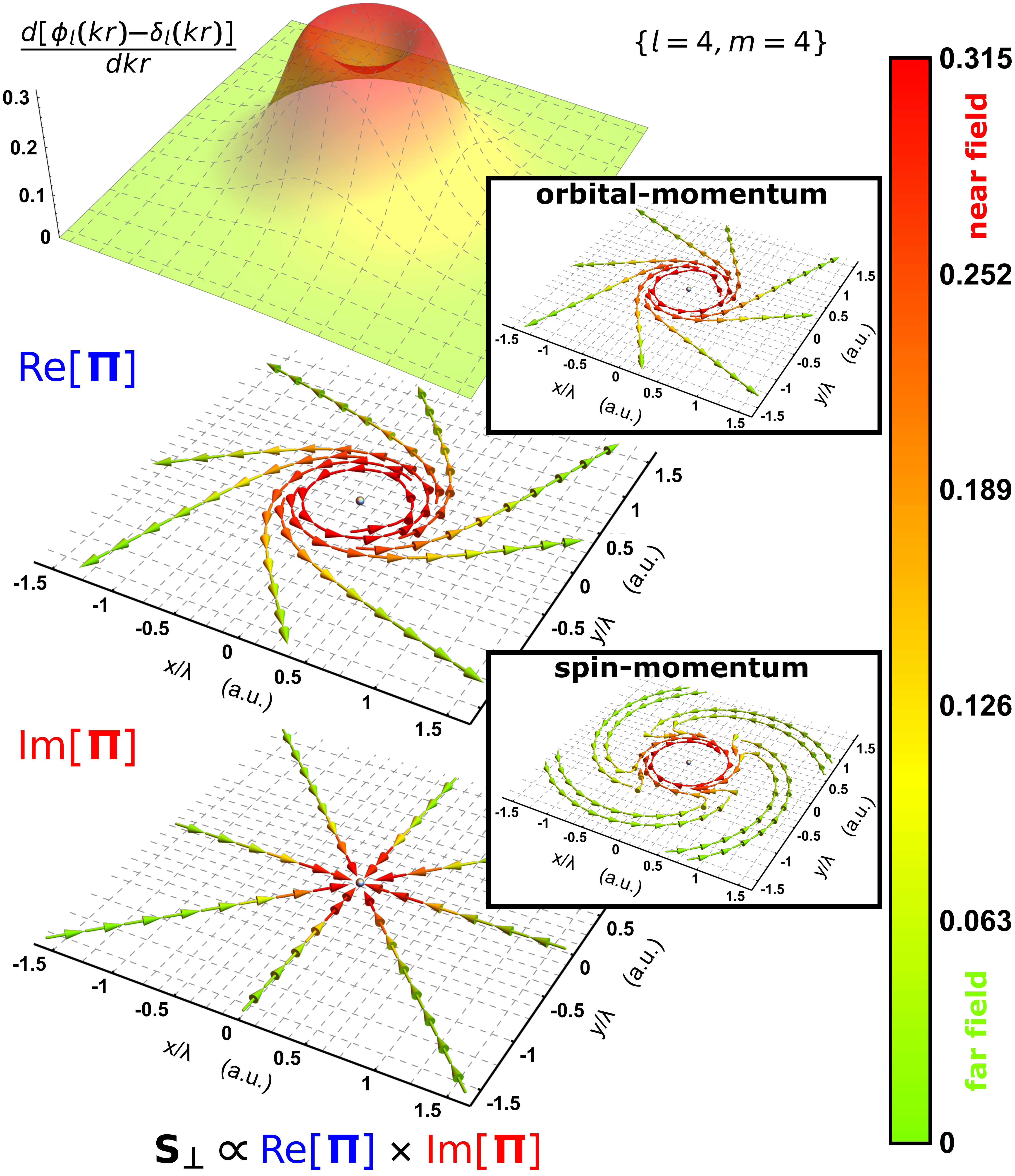} 
	\caption{Schematic representation of the near-field features for the ${\bf E}^{\rm TM}_{4,4}$-mode. From the SOI-term, the near- and the far-field region can be characterized in terms of the relative phase as $d\left[\phi_l-\delta_l\right]/dkr$. In the near-field region the spatial distribution of the multipole field exhibits a complex shape. This manifests itself by means of a rich and strikingly interesting structure of the local dynamical properties such as the complex Poynting vector, the orbital- and the spin-momentum given in Eqs. \eqref{poynting}--\eqref{pspin} (cf. Ref. \cite{Bekshaev2015} for further details on these properties). On the other side, in the far-field limit the field distribution tends to be orthogonal to the direction of propagation.}
	\label{fig3}
\end{figure}

In the following, neglecting the angular distribution, we will show how amplitude (leading to the SoP's modulation) and phase are intrinsically coupled together in the near-field region of propagating spherical waves. To elucidate this effect we start by writing the spherical Bessel functions from the recursive Rayleigh's formulas:
\begin{eqnarray}
\nonumber j_l(kr)&=&\left(-kr\right)^l\left[\frac{1}{kr}\frac{d}{dkr}\right]^l\left(\frac{\sin{kr}}{kr}\right)\\
&=&\frac{1}{kr}\left[P_l(kr)\cos{kr}+Q_l(kr)\sin{kr}\right],\\
\nonumber y_l(kr)&=&-\left(-kr\right)^l\left[\frac{1}{kr}\frac{d}{dkr}\right]^l\left(\frac{\cos{kr}}{kr}\right)\\
&=&\frac{1}{kr}\left[P_l(kr)\sin{kr}-Q_l(kr)\cos{kr}\right],
\end{eqnarray}
being $P_l(kr)$ and $Q_l(kr)$ real-valued polynomials of degree $l$-dependent. To simplify the analysis, we define the function $F_l(kr)\equiv P_l(kr)+ iQ_l(kr)=R_l(kr) e^{i\phi_l(kr)}$, from which the spherical Hankel functions associated to the propagating waves are given by
\begin{equation}
h_l^{(+)}(kr)=\frac{R_l(kr)}{kr} \exp{\left\{i[kr-\phi_l(kr)]\right\}}=\left[h_l^{(-)}\right]^*,
\end{equation}
where the asterisk denotes complex conjugation. From the latter expression we can observe that, despite the inhomogeneous spatial distribution stemming from the phase $\phi_l$, the spherical Hankel function behaves locally as a plane wave, i.e., is expressible as a product of an amplitude multiplied by a phase factor. Hence, analogously to Eq. \eqref{Eseparable}, we can say that the scalar function $h_l^{(\pm)}$ retains the spin-orbit factorizability (or separability) condition. This result applies both to the radial dependent functional coefficients $E_l^{\rm{(R)}}$ and $E_l^{(\Phi)}$. However, in Eq. \eqref{coeffb} one can readily see that $E_l^{(\Theta)}$ involves the first derivative with respect to $kr$, thus yielding the appearance of a relative phase. Indeed, since $F_l'(kr)= P'_l(kr)+ iQ'_l(kr)=(R'_l+iR_l\phi_l')e^{i\phi_l}=\tilde{R}_l(kr) e^{i\delta_l(kr)}$, the Eq. \eqref{coeffb} can be rewritten in the following form:
\begin{equation}
\frac{\left(krh_l^{(\pm)}(kr)\right)'}{kr}=\pm ih_l^{(\pm)}(kr)+\Delta^{(\pm)}_l(kr),
\label{separation}
\end{equation}
where we have defined
\begin{equation}
\Delta^{(\pm)}_l(kr)\equiv\frac{\tilde{R}_l(kr)}{kr}\exp{\left\{\pm i [kr-\delta_l(kr)]\right\}}. %e^{\pm i\left(kr-\delta_l(kr)\right)},
\label{SOI2}
\end{equation}
This $l$-dependent term (hereafter referred to as SOI-term) entails the nonseparability of the spin-orbit degrees of freedom in multipole fields, and therefore provides a suitable benchmark for claiming the fundamental emergence of intrinsic SOI-based effects at the nanoscale [see Fig. \ref{fig2}]. In fact, by a straightforward calculation it can be found that the relative phase $\phi_l-\delta_l$ influences dynamically only in the near-field region, thereby precluding the amplitude-phase separability [see inset of Fig. \ref{fig2}]. Moreover, since the amplitude $\tilde{R}_l$ vanishes in the far-field limit, $\Delta_l^{(\pm)}\to 0$, and the factorizability condition is recovered, leading, as expected, to the separable plane-like wave behavior. As displayed in the upper panel of Fig. \ref{fig3}, by gathering these features, the near and the far-field regions can be simply and accurately defined via the SOI-term as $d\left[\phi_l-\delta_l\right]/dkr$. In this way, we can set the far-field as the region where the relative phase is constant with respect to the dimensionless variable $kr$, namely, $\left[\phi_l-\delta_l\right]'\to 0$. On the other hand, according to our results, the near-field zone is strongly depending on the azimuthal mode order $l$, and is characterized by $\left[\phi_l-\delta_l\right]'\to 1/l$. It is important to highlight that, irrespective of the spatial intensity  distribution, the SOI-term tends to zero as $kr\to\infty$, thus confirming the, up to now assumed, intrinsic subwavelength character of SOI. This would enable the enhancement of SOI-based effects directly by raising the light intensity, still preserving the region wherein they appear.

From the above discussion it is worth noticing that the spin-orbit separability condition resembles the genuine concept of nonlocal quantum entanglement \cite{Pereira2014}. In fact, this mathematical structure describing the nonseparability between different degrees of freedom in a single physical system (SoP and phase in this case) is often termed as classical entanglement or correlation \cite{Spreeuw1998}. These correlations have already allowed to find interesting experimental capabilities for the realization of encoding and processing of polarization-dependent classical information (see Ref. \cite{Aiello2015} and references therein). 

\subsection{Definition of the near-field region based on the factorizability condition}
The distinction between the near- and the far-field region is often useful in theory of radiating systems because it provides a significant simplification in the analysis of the fields. Therefore, it would be convenient to have a precise condition to identify them accurately. In the particular case of an oscillating electric dipole,
\begin{eqnarray}
\nonumber{\bf E}_{\rm dipole}&=&\frac{k^2}{4\pi\varepsilon_0}\bigg[\left({\bf e}_r\times{\bf p}\right)\times{\bf e}_r\\
&+&\left(3\left({\bf e}_r\cdot{\bf p}\right){\bf e}_r-{\bf p}\right)\left(\frac{1}{k^2r^2}-\frac{i}{kr}\right)\bigg]\frac{e^{ikr}}{r},
\label{dipole}
\end{eqnarray}
the near- and the far-field terms are those proportional to $1/r^3$ and $1/r$, respectively \cite{Jackson}. In addition, the term proportional to $1/r^2$ is associated with the so-called intermediate-field or induction zone. These regions are actually characterized by a reasonable but arbitrary dependence with respect to the distance from the source $r$, assuming it as a emitter whose characteristic dimension $d$ is much smaller than the wavelength $\lambda$ and the distance $r$. This arbitrariness is even more evident for higher-order multipoles. Nonetheless, it can be demonstrated that the transverse component of the near-field term of the electric dipole given in Eq. \eqref{dipole} is closely related to the SOI-term of the corresponding multipole field. Besides giving a more accurate definition for the above regions in terms of the spin-orbit separability condition, our approach allows us to show that the term hindering the factorization solely influences in the near-field region. Therefore, this example provides a perfect test for demonstrating the agreement with the already existing theory, thus showing the universality of the optical SOI as a phenomenon occurring at the subwavelength scale. Furthermore, following Ref. \cite{Aiello2014}, we can find a subtle relationship between the near-field distribution given in Eq. \eqref{dipole} and the cross-polarization of a propagating beam described within the paraxial approximation. This may be the reason why the occurrence of SOI has been mostly identified under these distinct approaches. Indeed, as we have already seen, in the near-field region, the electric dipole cannot be generally expressed in a factorized form, and then the polarization and the propagation are mutually influenced. This behavior is similar to that of a propagating beam in inhomogeneous media \cite{Liberman1992}, and is the ultimate responsible for the appearance of SOI-based effects.

\subsection{Local dynamical properties of multipole fields: Poynting vector, spin- and orbital-momentum}
Until now, our analysis has been mainly focused into the influence of the SOI-term on the evolution of the SoP at the nanoscale. Still, according to the fundamental definition of the optical SOI, the propagation process must also be affected. Below we will address this remaining issue qualitatively by showing the behavior of the local momentum densities in the near-field zone.

It is well-known that in the simplest case of homogeneous plane-like waves the electromagnetic propagation is dictated by the real part of the complex Poynting vector \cite{Jackson},
\begin{figure*}[t!]
	\includegraphics[width=0.95\linewidth]{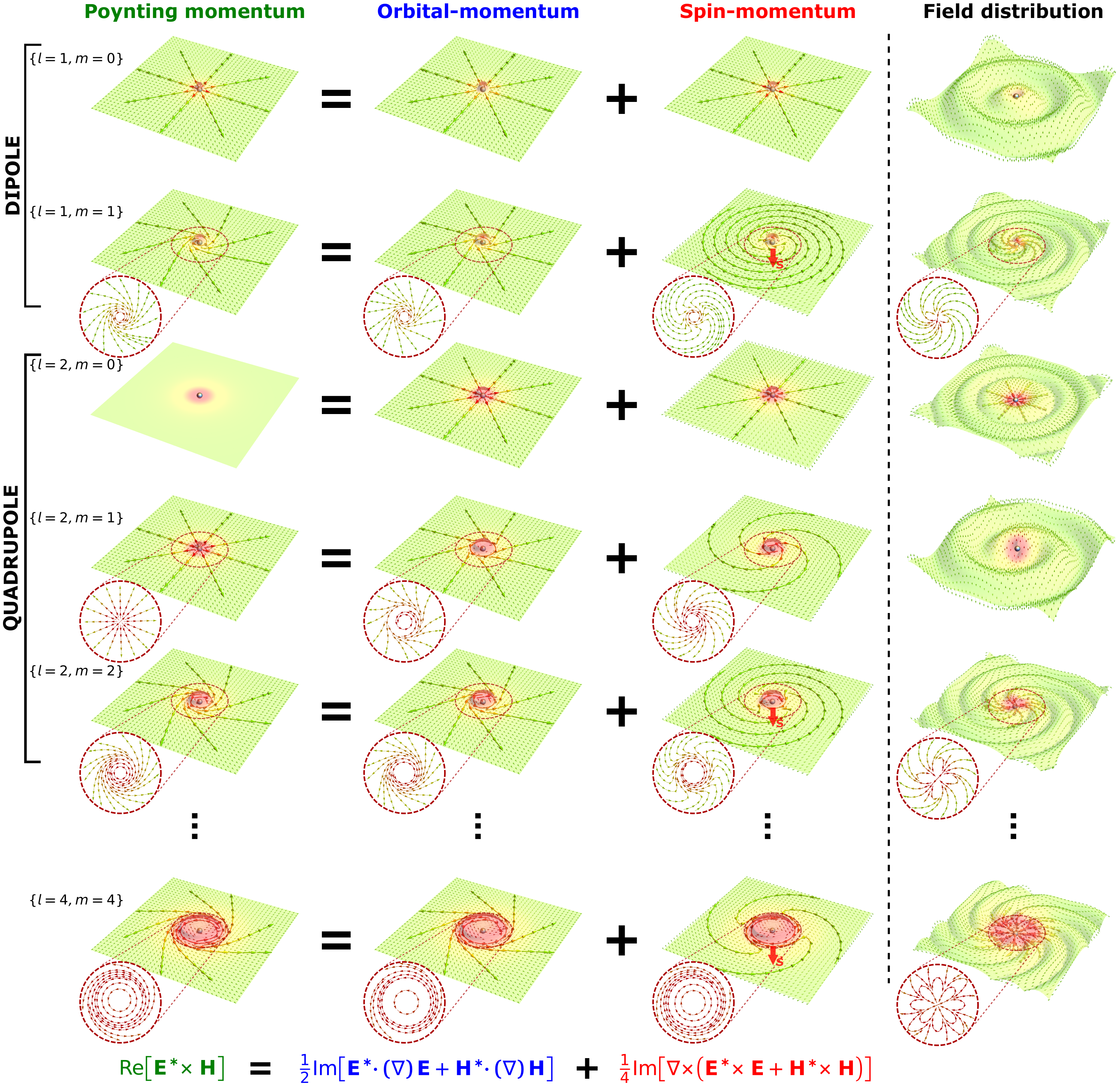} 
	\caption{Densities of the local dynamical properties for different multipole fields of $(l,m)$-order over the $xy$-plane. Columns 1 to 4 show the real part of the Poynting vector [Eq. \eqref{poynting}], the orbital- and spin-momentum [Eqs. \eqref{porbit} and \eqref{pspin}], and the locally normalized electric field distribution, respectively. The color coding used indicates the transition from the near- to the far-field zone, according to the scale represented in Fig. \ref{fig3}, for the corresponding value of $l$. Owing to the complex-shaped spatial field structure of the modes in-plane polarized, the local dynamical properties present streamlines that are sharply twisted in the near-field region. In fact, despite their seemingly planar character, there arises a transverse SAM. This is the responsible for an abrupt switching on the handedness of the spin-momentum near the source. Insets show a zoom-in view of this feature. }
	\label{fig4}
\end{figure*}
\begin{figure*}[t!]
	\includegraphics[width=0.95\linewidth]{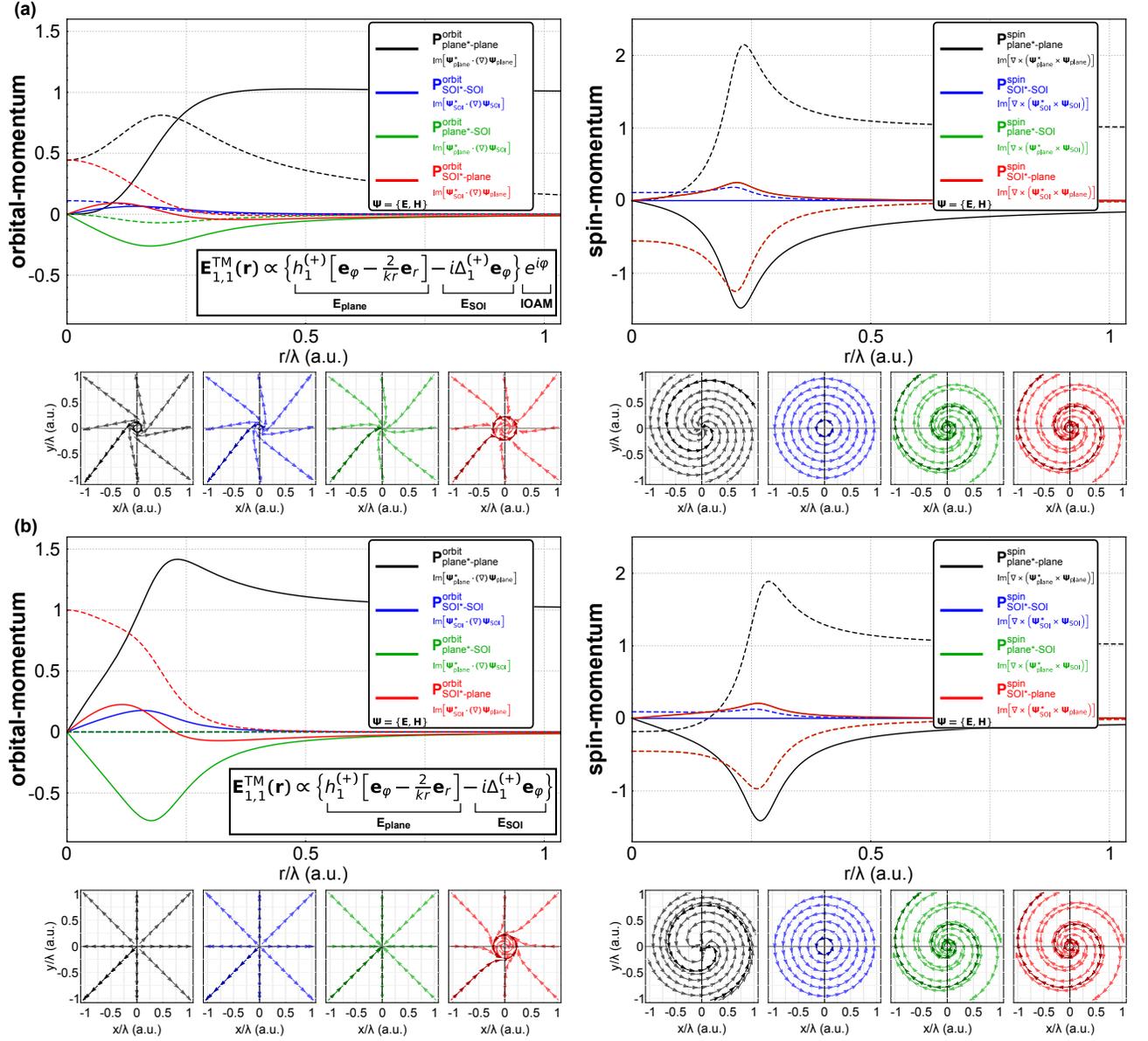} 
	\caption{Separate contributions to the local orbital- and spin-momentum densities for the ${\bf E}^{\rm TM}_{1,1}$-mode. Solid and dashed curves in the graphs show, respectively, the evolution of the radial and the azimuthal components normalized with respect to the corresponding local momentum density. Lower panels below the plots provides a better visual representation displaying the trajectories associated to the corresponding curves. In panel (a) the orbital- and spin-momentum are calculated by considering the whole VSWF, i.e., including both the SOI-term and the azimuthal angular dependence given by the phase term $e^{il\varphi}$. In panel (b) the underlying influence of the SOI-term is revealed by removing the azimuthal dependence. In this latter case, the plane-like contribution to the orbital-momentum density shows a purely radial behavior. The deviation from the radial direction is due to the nonseparability of the spin-orbit degrees of freedom. }
	\label{fig5}
\end{figure*}
\begin{equation}
{\bf \Pi}={\bf E}^*\times{\bf H},
\label{poynting}
\end{equation}
which points in the same direction as the wavevector ${\bf k}$. However, in structured optical fields, it is more convenient to decompose the latter quantity into the orbital (or canonical) and the spin contributions, $\text{Re}{\left[{\bf \Pi}\right]}={\bf P}^{\rm orbit}+{\bf P}^{\rm spin}$:
\begin{eqnarray}
{\bf P}^{\rm orbit}&=&\text{Im}{\left[{\bf E}^*\cdot\left(\nabla\right){\bf E}+{\bf H}^*\cdot\left(\nabla\right){\bf H}\right]}/2,
\label{porbit}\\
{\bf P}^{\rm spin\,}&=&\text{Im}{\left[\nabla\times\left({\bf E}^*\times{\bf E}\right)+\nabla\times\left({\bf H}^*\times{\bf H}\right)\right]}/4,
\label{pspin}
\end{eqnarray}
where we have used the notation ${\bf A}\cdot\left(\nabla\right){\bf B}=\sum_i{A_i\nabla B_i}$, and the proportionality factors have been absorbed into the normalization of the fields. Taking into account this separation, it has been recognized that the energy transport, characterizing the wave propagation, is associated with the orbital contribution to the total momentum of light (i.e., with its local phase gradient). On the other hand, the solenoidal-like spin-momentum has often been considered as a virtual divergence-less current. A deeper understanding of the role played by these properties deserves further efforts beyond the scope of this work (cf. Refs. \cite{Bekshaev2014,Bekshaev2015}). Despite that, by analyzing the influence of the SOI-term given in Eq. \eqref{SOI2}, we can show a number of dynamical characteristics underlying the occurrence of intrinsic SOI at the nanoscale. Specifically, as shown in Fig. \ref{fig4}, the most complex spatial field distribution arises for the multipole fields of $(l,l)$-order, polarized over the $xy$-plane. For these modes, the streamlines describing the spin and orbital energy flows are sharply twisted in the near-field region, thus showing a vortex-like behavior [see also Fig. \ref{fig5}]. Importantly, as also shown in Fig. \ref{fig3}, the spin-momentum abruptly switches its handedness. It can be demonstrated that this intriguing feature, together with the appearance of a transverse SAM, relies on the nonseparability of the spin-orbit degrees of freedom, thereby providing a clear signature of the emergence of intrinsic SOI. Notice that the existence of the spin-momentum does not depend on the SAM, i.e., there are modes without SAM over the $xy$-plane, still with a spin-momentum contribution. However, the change on the handedness of the spin-momentum only occurs for those modes with transverse SAM. In any case, the overall structure of the complex Poynting vector involves the joint action of the two types of momentum and is strongly dependent on the factorizability condition encompassed by the SOI-term. Indeed, we can show its effect on the light propagation, by plotting separately the radial and the azimuthal components for both the orbital- and spin-momentum densities [see Fig. \ref{fig5}]. Although, we have only considered the ${\bf E}^{\rm TM}_{1,1}$-mode, corresponding to a circularly polarized oscillating electric dipole, the present discussion is extensible to any other higher-order multipole field. In Fig. \ref{fig5}(a) we first consider the whole VSWF, i.e., including the SOI-term, the azimuthal angular dependence given by the phase $e^{il\varphi}$, and the plane-like part of the wave. For this case, owing to the presence of intrinsic-OAM, we can observe that all the contributions to the orbital-momentum are deviated from the radial direction. Therefore, in order to isolate the effect of the SOI-term we should remove, by hand, the azimuthal dependence. By doing so [see Fig. \ref{fig5}(b)], we find that the plane-like contribution to the orbital momentum is radially orientated, just as expected. Furthermore, we can observe a variety of anomalous effects such as the backward flow or the superluminal propagation. It has been established that these features are closely related to vortices and evanescent waves \cite{Bliokh2013b}. However, our approach is able to demonstrate that these effects are actually characteristics of intrinsic SOI.

As a final remark, it should be noted that spin-momentum locking has been demonstrated to be an inherent property of evanescent waves \cite{Mechelen2016,Fortuno2013}. This behavior, regarded as a manifestation of the quantum spin Hall effect of light \cite{Bliokh2015b}, is tied to the occurrence of SOI. Indeed, due to the transversality condition of the electromagnetic fields, $\nabla\cdot{\bf E}={\bf k}\cdot{\bf E}=0$, it was demonstrated that the transverse SAM and the wavevector are coupled to each other in such a way that ${\bf S}_\perp^{\rm evan}=\left(\text{Re}{\left[{\bf k}\right]}\times \text{Im}{\left[{\bf k}\right]}\right)/\text{Re}{\left[{\bf k}\right]}^2$. Remarkably, we can find a similar relationship between the complex Poynting vector and the transverse SAM for propagating waves:
\begin{equation}
{\bf S}_\perp^{\rm prop}=\pm\frac{\text{Re}{\left[{\bf \Pi}_{\rm (TE/TM)}\right]}\times\text{Im}{\left[{\bf \Pi}_{\rm (TE/TM)}\right]}}{W_{\rm (E/H)}},
\end{equation}
where $W_{\Psi}=\left|{\bf \Psi}^*\cdot{\bf \Psi}\right|$. It can be demonstrated that the validity of this result also relies on the nonseparability of the spin-orbit degrees of freedom, and then, it can be seen as a consequence of intrinsic SOI of light as well.

% % % % % % % % % % % % % % % % % % % % % % % % % % % % % % % % % % % % % % % % % % % % % % % % % % % % % % % % % % % % % % % % % % % % % % % %
% CONCLUSIONS
% % % % % % % % % % % % % % % % % % % % % % % % % % % % % % % % % % % % % % % % % % % % % % % % % % % % % % % % % % % % % % % % % % % % % % % %
\section{\label{sec:conclusion}Conclusions}
In summary, building on the already existing theory around the intrinsic SOI of light, we have put forward a suitable theoretical framework able to explain analytically its main features. The use of full-vector analysis involving spherical vector waves, highly appropriate for studying electromagnetic interaction at the nanoscale, allows us to obtain a factorizability condition for the electric (or magnetic) field that is only fulfilled in the far-field limit. In contrast, in the near-field region, both spin and orbit degrees of freedom get inherently coupled. It is important to remark that the nonseparability of the spin-orbit degrees of freedom, together with the transversality condition, are certainly the most important ingredients in order to unveil the mechanism leading to the classical emergence of the intrinsic SOI of light at the nanoscale. Even though the occurrence of SOI has already been theoretically reported in previous works (see, e.g., Refs. \cite{Liberman1992} and \cite{Bliokh2009}), there, the treatment was based on a perturbative analysis where the nonseparability between the spin-orbit degrees of freedom arose from higher-order terms stemming from the paraxial approximation. Importantly, in those demonstrations light is assumed to propagate as point-like particles, obeying Hamiltonian (or Lagrangian) dynamics and thereby neglecting its wave-like nature. Our finding, however, has the advantage of describing SOI of light from an analytical full-wave approach, providing a fundamental insight into the appearance of SOI-based effects in nano-optics. In spite of the simplicity of our treatment, it meets the overall prescriptions underlying the occurrence of optical SOI, showing that it naturally arises from the fundamental spin properties of Maxwell's equations and that necessarily appears at subwavelength distances. Furthermore, by using the spin-orbit factorizability condition, we can find a more accurate definition for the near-field region, thus removing the arbitrariness related to the dependence on the distance from the source. In view of the growing current interest in the optical SOI, we hope this analysis can be useful for the development of further optimum applications of SOI in classical \cite{Espinosa2017} and quantum nanophotonic devices \cite{Feber2015,Sollner2015}.

% % % % % % % % % % % % % % % % % % % % % % % % % % % % % % % % % % % % % % % % % % % % % % % % % % % % % % % % % % % % % % % % % % % % % % % %
% ACKNOWLEDGEMENTS
% % % % % % % % % % % % % % % % % % % % % % % % % % % % % % % % % % % % % % % % % % % % % % % % % % % % % % % % % % % % % % % % % % % % % % % %
\begin{acknowledgments}
The authors are grateful to F.J. Rodr\'iguez-Fortu\~no for valuable comments and discussions. This work was supported by funding from contract TEC2014-51902- C2-1-R (MINECO/FEDER, UE).
\end{acknowledgments}

% % % % % % % % % % % % % % % % % % % % % % % % % % % % % % % % % % % % % % % % % % % % % % % % % % % % % % % % % % % % % % % % % % % % % % % %
% REFERENCES
% % % % % % % % % % % % % % % % % % % % % % % % % % % % % % % % % % % % % % % % % % % % % % % % % % % % % % % % % % % % % % % % % % % % % % % %
\end{document}